The Potential Impact of Increased Hospital Capacity to Contain and Control Ebola in Liberia

Eric T. Lofgren[1], Caitlin M. Rivers[1], Madhav V. Marathe[1], Stephen G. Eubank[1], Bryan L. Lewis[1]

[1]Network Dynamics and Simulation Science Lab, Virginia Bioinformatics Institute, Virginia Tech, Blacksburg, VA USA

Corresponding Author:
Eric T. Lofgren, Ph.D.
Network Dynamics and Simulation Science Laboratory
Virginia Bioinformatics Institute
Virginia Tech
1880 Pratt Drive (0477)
Blacksburg, VA 24061 USA
Tel: 540-231-1349
Email: lofgrene@vbi.vt.edu

**Abstract:**

West Africa is currently experiencing a severe outbreak of Ebola virus disease (EVD). As part of the international effort to address this outbreak, the United States has committed to building specialized Ebola treatment facilities with 1700 beds. However, the effectiveness of this increase in the available healthcare facilities to treat Ebola is unclear, especially in light of the rapidly increasing number of cases. Adapting a previously validated mathematical model of Ebola in West Africa, we examine the potential impact of an increase in hospital capacity to mitigate the impact of Ebola under several scenarios, ranging from the planned scenario of 1700 beds in 10 weeks to a considerably more aggressive approach of twice the number of beds in 5 weeks. We find that even for the most aggressive scenarios, while increasing the availability of healthcare reduces the number of Ebola cases and slows the outbreak, it is not sufficient to stop the epidemic within the next three months. We find that only a combination of increased hospital beds and a dramatic decrease in the rate of transmission within the community can bring the epidemic under control within the near future.

**Introduction**

Since March of 2014, West Africa has been experiencing an outbreak of Ebola virus disease (EVD) of unparalleled size, beginning in Guinea and spreading to Liberia and Sierra Leone, with sporadic cases in Nigeria, Senegal and outside Africa. Model-based projections of the progress of the outbreak forecast an increasingly dire epidemic(1-3), with no suggestion that it will come under control in the foreseeable future without a robust response from the international community.

As part of the response to the epidemic, the United States has committed to constructing up to 17 specialized Ebola treatment centers with 1700 beds(4). These centers are intended to provide state-of-the-art care for patients with EVD, with appropriate equipment and supplies to follow infection control guidelines, and the means to safely handle the remains of those who die from the disease. Although beneficial to patients who are treated in these facilities, the impact of their introduction on the epidemic as a whole – especially as the number of new infections continues to grow - remains unknown.

Mathematical models may be used to forecast the effect of an intervention before it is put into practice, building off what is presently known about the epidemic in order to make predictions. In this study we adapt a previously validated model of the current Ebola outbreak in Liberia to examine under what circumstances the construction of specialized Ebola treatment centers might impact the epidemic, and what additional measures may need to be applied to bring it fully under control in the near future.

**Methods:**

Case Time series

A time series of laboratory confirmed, suspected and probable cases of EVD as reported by the World Health Organization and the Liberian Ministry of Health was collected, and is available in curated form at https://github.com/cmrivers/ebola. A combination of all three types of cases is thought, given the severe constraints on laboratory testing within Liberia, to better represent the current state of the outbreak.

Mathematical Model

A two-stage compartmental model was used to represent the course of the outbreak up to the present, and the potential impact of increasing the capacity of the healthcare system within the near future. In the first stage, the compartmental structure is similar to that in Legrand et al.(5), which has been previously used to describe the 1995 outbreak in the Democratic Republic of the Congo, the 2000 outbreak in Uganda, and the current outbreak in both Liberia and Sierra Leone(1). Technical details of the model can be found in (1). Briefly, the population is divided into six compartments, as shown in Figure 1A. Susceptible individuals without EVD may become exposed (E) after contact with an infected individual, and in turn become infectious (I) once the disease's incubation period has passed. A proportion of these individuals may seek medical treatment (H). Whether or not they seek treatment, infected patients may experience one of two possible outcomes. Either they die of their infection, wherein they may infect others during the burial process (F) before they are removed from the model (R), or they may recover, whereupon they are also removed.

The second stage of the model, which contains more detail regarding improvements to the healthcare system in Liberia, is shown in Figure 1B. Here, the H compartment used in the first stage is broken up into three separate compartments: $H_E$, $H_C$ and $H_H$ which represent treatment at specialized Ebola Treatment Centers (ETCs), local Ebola Community Centers (ECCs) and home-based medical care using take-home kits, respectively. Each of these levels of care also has a corresponding funeral class ($F_E$, $F_C$ and $F_H$) to model the burial of those patients who die of their infection while under the supervision of the healthcare system. The rate at which patients are admitted into each type of healthcare is based on the number of available beds – when capacity is high, patients are admitted more frequently to ETCs, while when capacity is low they are more likely to be admitted to ECCs or given take-home kits during times of extreme bed shortages. The equations that govern this stage of the model are shown in Eq. 1, and the parameters used in Table 1.

Model Validation and Stochastic Simulation

The first stage of the model is fit and validated against current outbreak data using a weighted least-squares optimization, as discussed in Rivers et al., 2014(1). This model provides a mathematical representation of the epidemic up to Oct. 1, 2014. After this date, the second stage of the model is stochastically simulated using Gillespie's algorithm with a tau-leaping approximation(6, 7) using the same parameters as the validated model fit as described above, and with initial populations in each compartments matching the states of the deterministic compartments on Oct. 1, with patients in the H compartment assumed to be members of $H_C$. This model is simulated 250 times for a 90-day period, to give a collection of epidemic forecasts accounting for the role of random chance(8). All models were implemented using Python 2.7 and the StochPy library for stochastic simulation(9).

Modeled Scenarios

Four potential interventions were modeled, based off the goal set by the U.S. government of 1700 ETC beds rolled out in 10 weeks, along with three times that number of ECC beds. The first scenario models the planned intervention – the addition of 170 beds per week for 10 weeks ('planned rollout'). Additionally, we examine the deployment of the proposed number of beds in half the amount of time ('fast rollout'), the deployment of twice the number of beds in the proposed 10-week period ('double rollout'), and the deployment of twice the number of beds in half the amount of time ('super rollout'). Additionally, each scenario is considered under three different potential levels of efficacy for infection control and safe burial practices. In the "ideal" case, ETCs reduce the transmission parameters for both hospital and funeral transmission by 100%, ECCs by 70% and home kits by 50%, reducing their corresponding $β_H$ and $β_F$ parameters. In the "optimistic" case, this reduction is 90%, 50% and 10% respectively, and in the "conservative" case, this reduction is 80%, 20% and 0%. A baseline scenario, where the epidemic is allowed to continue without intervention, was also modeled (Table 2).

In addition to these scenarios, a number of potential supplemental or coexisting interventions were modeled. Using the most aggressive scenario – the super rollout with ideal infection control, we model the effect of a 25% and 50% reduction in time until hospitalization ($\Upsilon_H$), representing increased access to care due to the strategic placement of ETCs, or an increased willingness to seek care earlier during the course of infection due to public perception of an increased chance of survival. Finally, we combine the super rollout and 25% time to hospitalization decrease outlined above with a 25% or 50% reduction in community transmission ($\beta_I$) to represent co-occurring interventions in the community at large.

Human Subjects

As this study uses publically available data without personal identifiers, it was determined not to require IRB approval.

**Results**

Model Validation and Baseline Prediction

The first-stage deterministic model fit the reported cumulative cases well (Figure 2). The baseline scenario, with no additional control beyond what already exists in Liberia, predicts an extremely severe epidemic, with a median of 151,245 (interquartile range (IQR): 148,045 - 153,742) incident cases in the 90 days after Oct. 1 (Figure 3).

Any scenario involving an increase in available hospital capacity resulted in a dramatic decrease in the overall number of cases (Figure 4). The planned rollout of 1700 ETC beds in 10 weeks and three times that number of ECC beds resulted in a median of 67,134 (IQR: 65,647 – 69,034) incident cases using the ideal parameters, 82,911 (IQR: 80,535 – 84,902) using the optimistic parameters, and 105,699 (IQR: 103,202 – 108,175) using the conservative parameters. The faster rollout of the same number of beds in five weeks resulted in a median of 64,253 (IQR: 62,793 – 66,069), 77,818 (IQR: 75,995 – 79,740) and 97,922 (95,525 – 100,372) cases in using the ideal, optimistic and conservative parameters.

When the number of ETC and ECC beds was doubled, the predicted number of incident cases was 62,451 (IQR: 61,211 – 63,969) for the ideal parameters, 74,788 (IQR: 73,090 – 76,791) for the optimistic parameters, and 93,554 (IQR: 90,889 – 95,900) for the conservative parameters. When the number of beds was doubled and the period of time in which they were deployed was reduced to five weeks, the projected number of incident cases was 60,560 (IQR: 58,826 – 61,983) for the ideal parameters, 71,807 (IQR: 69,957 – 73,715) for the optimistic parameters and 87,897 (IQR: 86,157 – 89,730) for the conservative parameters. Importantly, while all of these scenarios predict a large decrease in the number of incident cases of EVD, none of them suggest that the epidemic is fully under control, with the cumulative number of cases of EVD still growing exponentially (Figure 5).

When combining the most effective healthcare-only intervention (super rollout using ideal parameters) with 25% and 50% reductions in time to hospitalization, the number of incident cases drops to a median of 38,966 (IQR: 37,898 – 39,974) for a 25% reduction and 20,383 (IQR: 19,922-20,907) for a 50% reduction. Additionally, the 50% reduction in time to hospitalization moves the epidemic from growing exponentially to growing linearly (Figure 6). Community-based interventions have similarly pronounced effects, with a 25% reduction in community transmission (accompanying the super rollout and a 25% decrease in time to hospitalization) resulting in a median predicted number of incident cases of 18,771 (IQR: 18,363 – 19,180) and a 50% reduction in community transmission resulting in a median of 10,263 (IQR: 10,077 – 10,486) incident cases. The 25% scenario also shows the epidemic having moved to a linear growth phase, and the 50% reduction results in a cumulative case curve that has begin to level off completely, indicating that the peak of infection has passed (Figure 7).

**Discussion**

The results of these forecasts suggest several key aspects to the proposed augmentation of the Liberian healthcare system with newly constructed hospitals meant to treat Ebola patients. The first is that these facilities have a considerable impact on the overall burden of disease, resulting in over 45,000 fewer cases between the baseline, uncontrolled scenario and the next worst scenario with improved healthcare infrastructure. Despite this, an improved healthcare system alone cannot stop the current epidemic, nor even move it toward more manageable linear growth. These findings agree with those suggested by other modeling studies (10).

The second is that the variation within a given scenario (e.g. between the ideal and conservative parameters) is considerably higher than between scenarios (e.g. between the planned rollout and the super rollout), which highlights the importance of hospital infection control in preventing infections. Simply increasing the overall number of beds has less value in preventing infections than ensuring that those beds are well staffed and equipped, and capable of handling EVD patients without causing secondary transmission. This finding is mirrored by the small number of cases in the United States, where secondary transmission was the result of failures in infection control, rather than an absence of available facilities.

Finally, these results underscore the importance of the interaction between the community in Liberia and the healthcare system. Previous work has shown that even with extremely high levels of hospitalization, the epidemic can still continue based on the time infected patients spend in the community *before* seeking treatment. Even a relatively modest 25% decrease in the time until an infected patient seeks treatment has a dramatic impact on the course of the outbreak, and suggests that access to treatment, be it optimized placement of treatment centers, better follow-up of exposed contacts, or another factor may be beneficial in bringing the epidemic under control. Similarly, reductions in community transmission, when done in concert with improved treatment and reduced time to hospitalization, results in an epidemic that is only a fraction of the size of an control strategy relying strictly on improved healthcare infrastructure.

While improvements in healthcare infrastructure have the potential to markedly improve the ongoing Ebola epidemic in Liberia, they are not sufficient to bring the outbreak under control, even under fairly aggressive assumptions about the quantity and speed at which they are constructed. Instead, a multifaceted approach is needed. In conjunction with increases in the number of available beds, basic tools of public health, such as improved messaging, as well as there careful consideration of how patients get to the treatment centers once they are in place, will be essential. In any scenario, a long-term commitment to healthcare infrastructure, hospital infection control, and community-level interventions will be needed from the international community.

**Acknowledgements**

The authors would like to thank Katie Dunphy, Jesse Jeter, P. Alexander Telionis, James Schlitt, Jessie Gunterand Meredith Wilson for their assistance and support. The authors would also like to acknowledge the participants in the weekly briefings organized by DTRA, BARDA and NIH, including Dave Myer, Aiguo Wu, Mike Phillips, Ron Merris, Jerry Glashow, Dylan George, Irene Eckstrand, Kathy Alexander and Deena Disraelly for their comments and feedback. Finally, the authors would like to thank Timo Maarleveld for his assistance with the StochPy library.

**References**

1. Rivers CM, Lofgren ET, Marathe M, Eubank S, Lewis BL (2014) Modeling the Impact of Interventions on an Epidemic of Ebola in Sierra Leone and Liberia. *PLoS Currents Outbreaks*.

2. WHO Ebola Response Team (2014) Ebola Virus Disease in West Africa - The First 9 Months of the Epidemic and Forward Projections. *N Engl J Med*. 371: 1481-95

3. Meltzer MI et al. (2014) Estimating the future number of cases in the ebola epidemic --- liberia and sierra leone, 2014--2015. *MMWR Surveill Summ* 63:1–14.

4. Cooper H, Shear MD, Grady D (2014) U.S. to Commit Up to 3,000 Troops to Fight Ebola. *New York Times*. Available at: http://www.nytimes.com/2014/09/16/world/africa/obama-to-announce-expanded-effort-against-ebola.html?_r=0 [Accessed October 29, 2014].

5. Legrand J, Grais RF, Boelle PY (2007) Understanding the dynamics of Ebola epidemics. *Epidemiology and Infection*. 135: 610-21

6. Gillespie DT (1977) Exact stochastic simulation of coupled chemical reactions. *J Phys Chem* 81:2340–2361.

7.  Gillespie DT (2001) Approximate accelerated stochastic simulation of chemically reacting systems. *The Journal of Chemical Physics*. 115(4): 1716-33

8.  Lofgren ET (2012) Visualizing results from infection transmission models: a case against "confidence intervals". *Epidemiology* 23:738–741.

9.  Maarleveld TR, Olivier BG, Bruggeman FJ (2013) StochPy: A Comprehensive, User-Friendly Tool for Simulating Stochastic Biological Processes. *PLoS One* 8:e79345.

10. Lewnard JA et al. Dynamics and control of Ebola virus transmission in Montserrado, Liberia: a mathematical modelling analysis. *The Lancet Infectious Diseases*.

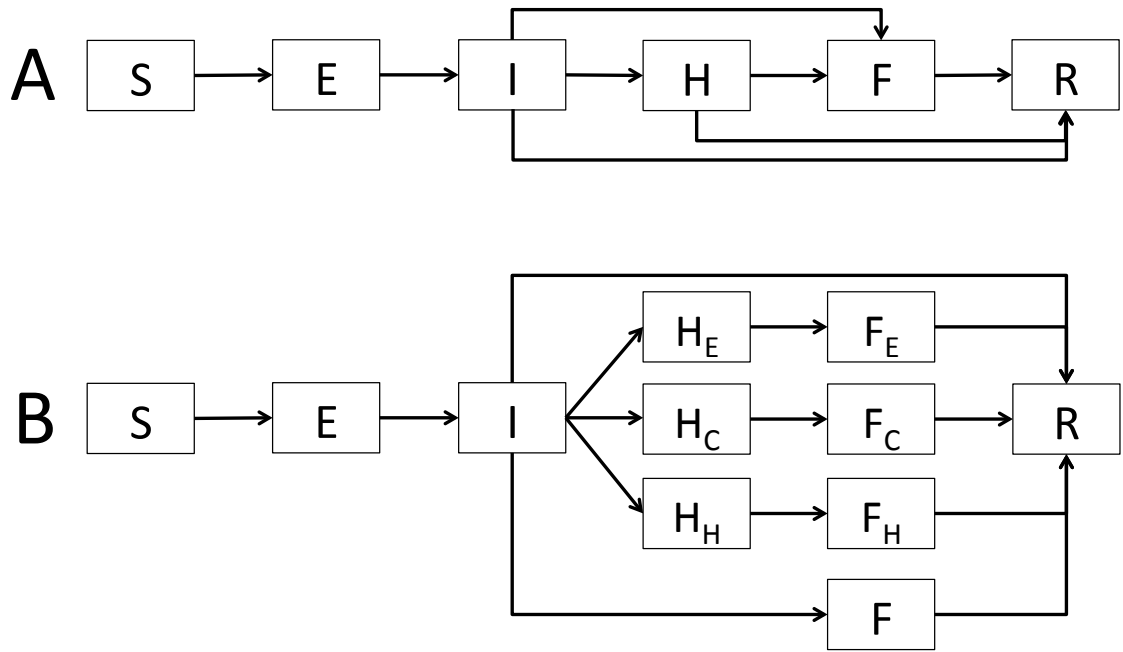

**Fig. 1. Compartmental flow of a two-stage mathematical model of the Ebola Epidemic in Liberia, 2014.** In the deterministic model of the epidemic up to Oct. 1 (Panel A) the population is divided up into six compartments: Susceptible (S), Exposed (E), Infectious (I), Hospitalized (H), Funeral (F) and Recovered/Removed (R). In the stochastic model of an improved healthcare infrastructure, the H and F categories are divided into compartments representing patients in specialized Ebola Treatment Centers ($X_E$), treated in community centers ($X_C$) or given take-home kits ($X_H$). Arrows indicate possible transitions between states.

**Eq. 1.**

$$\frac{dS}{dt} = -\frac{\beta_I SI + \beta_H \tau_E SH_E + \beta_H \tau_C SH_C + \beta_H \tau_H SH_H + \beta_F \tau_E SF_E + \beta_F \tau_C SF_C + \beta_F \tau_H SF_H + \beta_F SF}{N}$$

$$\frac{dE}{dt} = \frac{\beta_I SI + \beta_H \tau_E SH_E + \beta_H \tau_C SH_C + \beta_H \tau_H SH_H + \beta_F \tau_E SF_E + \beta_F \tau_C SF_C + \beta_F \tau_H SF_H + \beta_F SF}{N} - \alpha E$$

$$\frac{dI}{dt} = \alpha E - [\gamma_H \theta_1 (1 - \frac{H_E}{\kappa_E}) + \gamma_H \theta_1 (\frac{H_E}{\kappa_E})(1 - \frac{H_C}{\kappa_C}) + \gamma_H \theta_1 (\frac{H_E}{\kappa_E})(\frac{H_C}{\kappa_C}) + \gamma_I (1 - \theta)(1 - \delta_1) + \gamma_D (1 - \theta_1) \delta_1] I$$

$$\frac{dH_E}{dt} = \gamma_H \theta_1 (1 - \frac{H_E}{\kappa_E}) I - [\gamma_{DH} \delta_2 + \gamma_{IH}(1 - \delta_2)] H_E$$

$$\frac{dH_C}{dt} = \gamma_H \theta_1 (\frac{H_E}{\kappa_E})(1 - \frac{H_C}{\kappa_C}) I - [\gamma_{DH} \delta_2 + \gamma_{IH}(1 - \delta_2)] H_C$$

$$\frac{dH_H}{dt} = \gamma_H \theta_1 (\frac{H_E}{\kappa_E})(\frac{H_C}{\kappa_C}) I - [\gamma_{DH} \delta_2 + \gamma_{IH}(1 - \delta_2)] H_H$$

$$\frac{dF_E}{dt} = \gamma_{DH} \delta_2 H_E - \gamma_F F_E$$

$$\frac{dF_C}{dt} = \gamma_{DH} \delta_2 H_C - \gamma_F F_C$$

$$\frac{dF_H}{dt} = \gamma_{DH} \delta_2 H_H - \gamma_F F_H$$

$$\frac{dF}{dt} = \gamma_D (1 - \theta_1) \delta_1 I - \gamma_F F$$

$$\frac{dR}{dt} = \gamma_I (1 - \theta_1)(1 - \delta_1) I + \gamma_{IH}(1 - \delta_2) H_E + \gamma_{IH}(1 - \delta_2) H_C + \gamma_{IH}(1 - \delta_2) H_H$$

**Table 1. Model Parameters and Fitted Values for a Model of an Ebola Epidemic in Liberia with Healthcare Infrastructure Improvements, 2014.**

| Parameter | Value |
| --- | --- |
| Contact Rate, Community ($\beta_I$) | 0.160 |
| Contact Rate, Hospital ($\beta_H$) | 0.062 |
| Contact Rate, Funeral ($\beta_F$) | 0.489 |
| Incubation Period ($1/\alpha$) | 12 days |
| Time until Hospitalization ($1/\gamma_H$) | 3.24 days |
| Time from Hospitalization to Death ($1/\gamma_{DH}$) | 10.07 days |
| Duration of Traditional Funeral ($1/\gamma_F$) | 2.01 days |
| Duration of Infection ($1/\gamma_I$) | 15.00 days |
| Time from Infection to Death ($1/\gamma_D$) | 13.31 days |
| Time from Hospitalization to Recovery ($1/\gamma_{IH}$) | 15.88 days |
| Fraction of infected hospitalized ($\theta_1$) | 0.20 |
| Case Fatality Rate, Unhospitalized ($\delta_1$) | 0.500 |
| Case Fatality Rate, Hospitalized ($\delta_2$) | 0.500 |
| Infection Control Efficacy ($\tau_X$) | Varies (Range: 0 – 1) |
| Available Hospital Beds ($\kappa_X$) | Varies (Range: 0 – 10200) |

**Table 2. Modeled Scenarios of an Ebola Epidemic in Liberia with Healthcare Infrastructure Improvements, 2014.**

| Name | ETC Beds | ECC Beds | Deployment Time |
| --- | --- | --- | --- |
| Planned | 1700 | 5100 | 10 weeks |
| Fast | 1700 | 5100 | 5 weeks |
| Double | 3400 | 10200 | 10 weeks |
| Super | 3400 | 10200 | 5 weeks |

**Fig. 2. Model Fit of a Compartmental Model of an Ebola Epidemic in Liberia, 2014.** Red dots depict the reported number of cumulative cases of Ebola, and the solid black line indicated the deterministic fit of the model to Oct. 1, 2014.

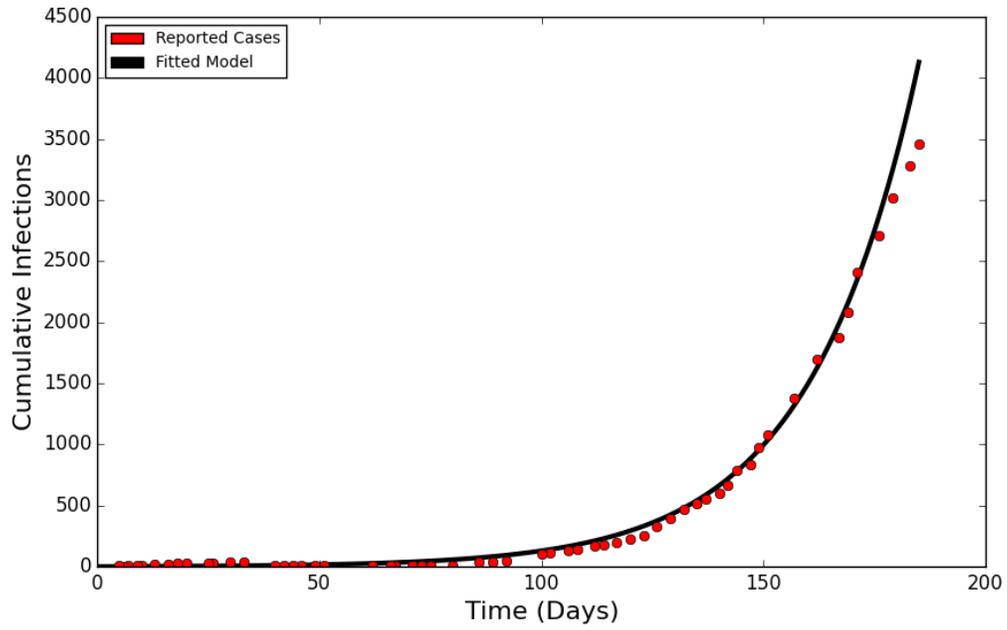

**Fig. 3. Stochastic Forecast of Uncontrolled Ebola Epidemic in Liberia, 2014.**
Solid black line indicates deterministic model fit up to Oct. 1, with the grey lines depicting 250 stochastic forecasts of the epidemic. Areas of denser color indicate larger numbers of forecasts with those values.

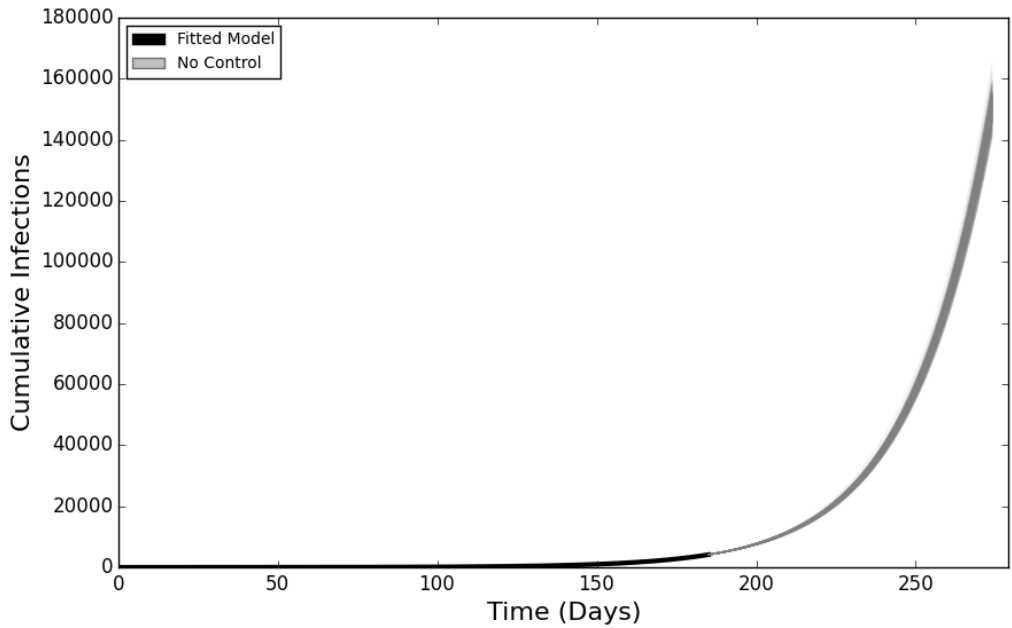

**Fig. 4. Distribution of Incident Ebola Cases in Liberia 90 Days after Oct. 1, 2014 with Improving Healthcare Infrastructure.** Each panel represents a particular scenario (along with the uncontrolled baseline) for all three parameter combinations. Box plots depict the median, interquartile range and 1.5 times the interquartile range for each scenario. Each simulated forecast is shown as a single point, jittered to show the complete distribution of the results.

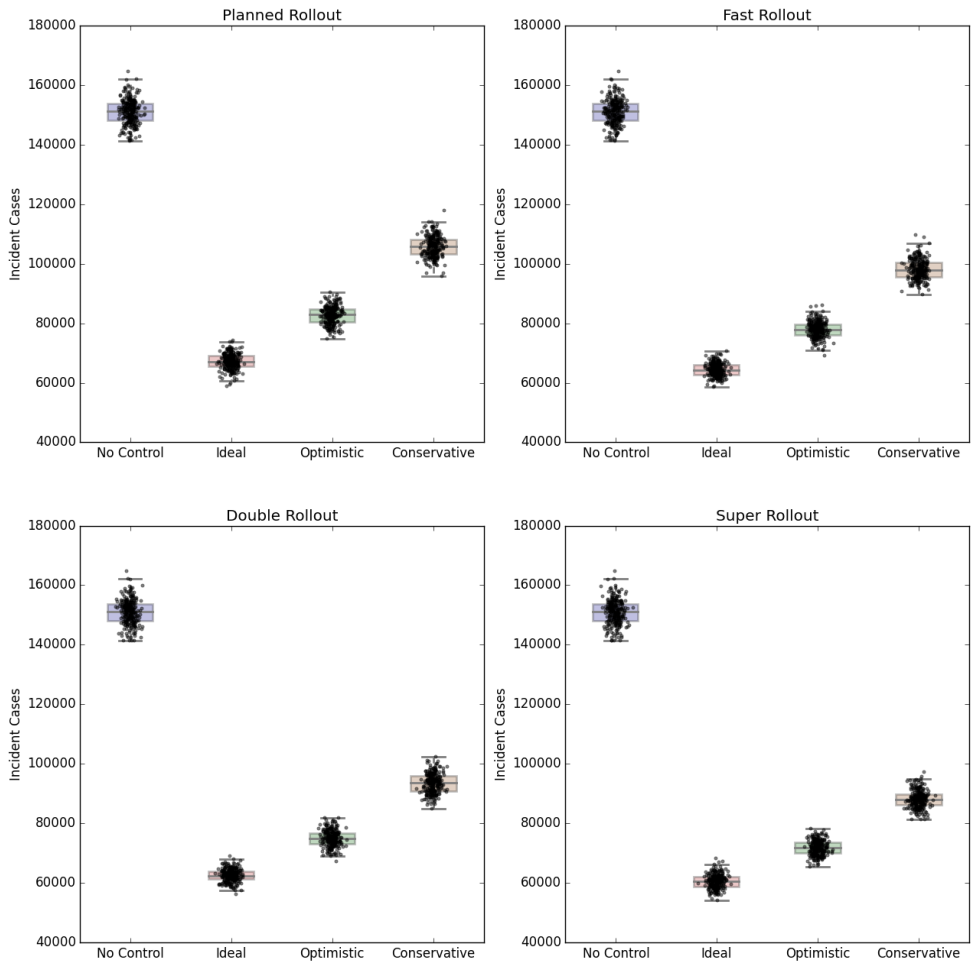

**Fig. 5 Cumulative Ebola Cases in Liberia 90 Days after Oct. 1, 2014 with Improving Healthcare Infrastructure.** Each panel represents a particular scenario (along with the uncontrolled baseline), with the solid black line indicating the deterministic model fit up to Oct. 1, 2014 with each colored line representing a single simulated forecast of the epidemic with ideal (blue), optimistic (green) or conservative (red) parameters. Areas of darker color indicate more forecasts with that result.

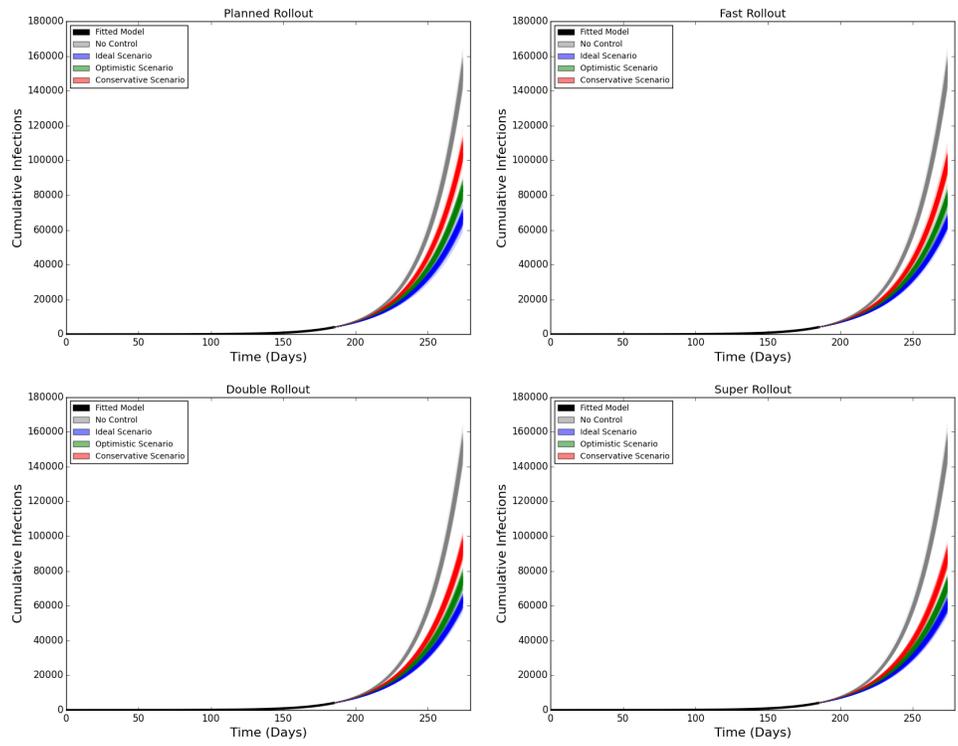

**Fig. 6 Cumulative Ebola Cases in Liberia 90 Days after Oct. 1, 2014 with Improving Healthcare Infrastructure and Reduced Time to Hospitalization.**
The solid black line indicated the deterministic model fit up to Oct. 1, 2014 with each colored line representing a single simulated forecast of the epidemic, either uncontrolled (grey), with the super rollout and no reduction in time to hospitalization (blue), the super rollout and a 25% reduction in time to hospitalization (green) or the super rollout and a 50% reduction in time to hospitalization (red). Areas of darker color indicate more forecasts with that result.

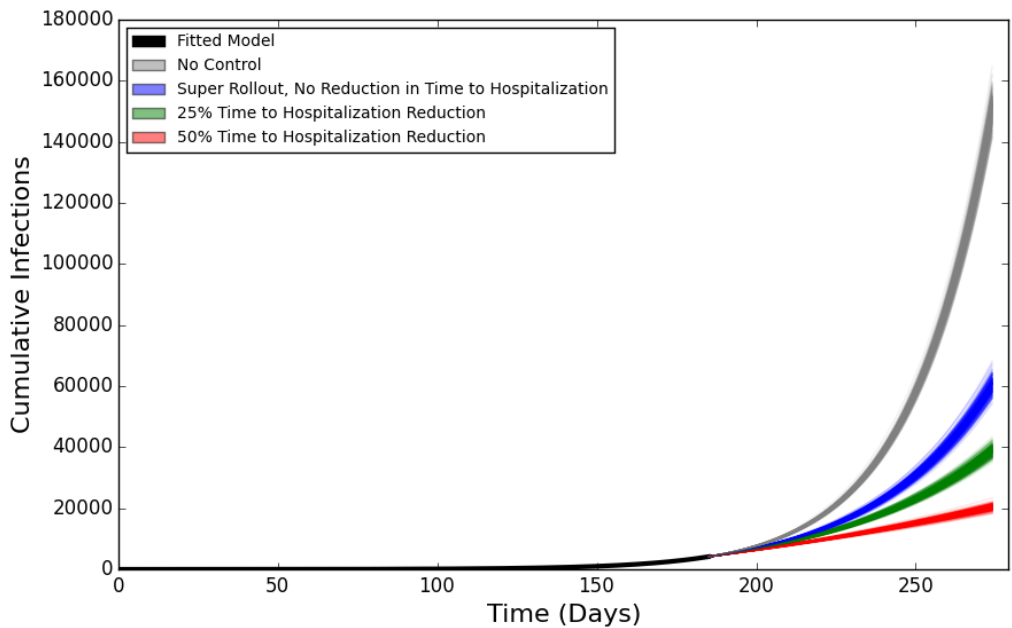

**Fig. 7 Cumulative Ebola Cases in Liberia 90 Days after Oct. 1, 2014 with Improving Healthcare Infrastructure, Reduced Time to Hospitalization and Reductions in Community Transmission.** The solid black line indicated the deterministic model fit up to Oct. 1, 2014 with each colored line representing a single simulated forecast of the epidemic, either the super rollout and a 25% reduction in time to hospitalization (grey) the super rollout and a 25% reduction in time to hospitalization and a 25% (blue) or 50% (green) reduction in community transmission. Areas of darker color indicate more forecasts with that result.

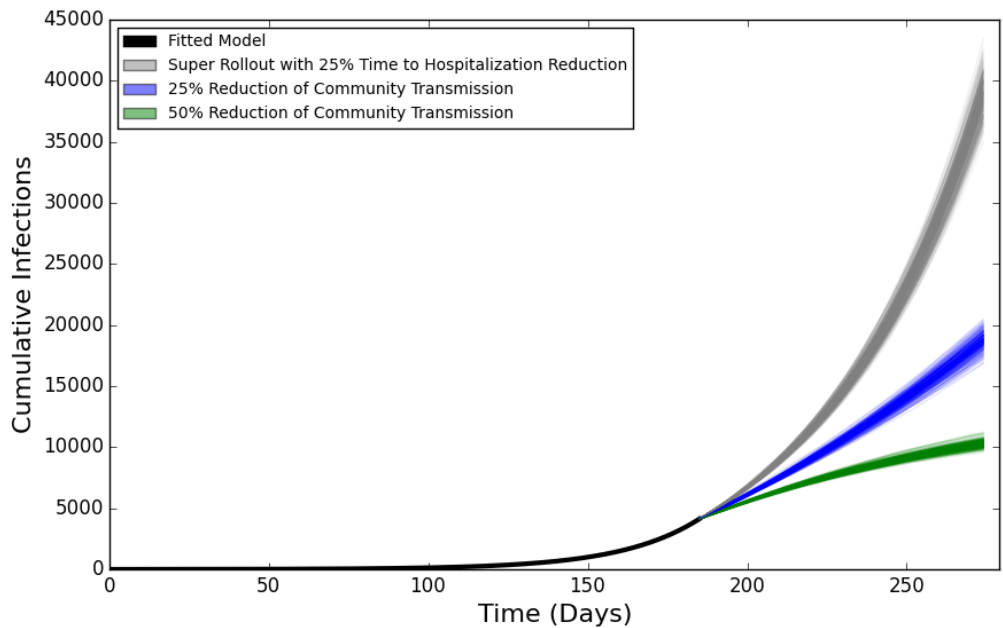